\documentclass[5p,final,twocolumn]{elsarticle}
\usepackage{amssymb}
\usepackage{epsfig}
\usepackage{amsmath}
\begin{document}

\begin{frontmatter}

\title{Quantum-classical hybrids in a simplified model of QED and geometric phase induced by charged particle trajectory}
\author{T. Koide}
\address{Instituto de F\'{\i}sica, Universidade Federal do Rio de Janeiro, C.P.
68528, 21941-972, Rio de Janeiro, Brazil }
\begin{abstract}
We derive a model of quantum-classical hybrids for a simplified model of quantum electrodynamics 
in the framework of the stochastic variational method.
In this model, charged particle trajectories are affected by the interaction with quantized electromagnetic fields, and 
this quantum-classical interaction induces a displacement current. 
We further investigate a geometric phase 
in the wave functional of the gauge field configuration, 
which is induced by adiabatic motions of the charged particles.
This phase contains the quantum-classical backreaction effect and usual 
Berry's phase is reproduced in the vanishing limit of 
the fluctuation of the charged particle trajectories.
\end{abstract}


\end{frontmatter}

\section{Introduction}

Quantum-classical hybrid (QCH) theory is intended to describe coexisting systems of classical and quantum 
degrees of freedom. 
There are several situations where the idea of QCH is applicable: quantum measurement \cite{sudar}, 
quantum-to-classical transition in early universe \cite{kiefer}, 
Einstein gravity interacting with quantum objects \cite{diosi}, Berry's phase in adiabatic processes \cite{zhang}, 
spin hydrodynamics \cite{koidesemh} and so on. 
Moreover, QCH has been studied to simplify complex numerical simulations in quantum chemistry \cite{bur}. 

There are various proposals for QCH, but it is not easy to maintain consistency requirements 
such as the energy conservation, the positivity of probability and Ehrenfest's theorem, which are summarized in Ref.\ \cite{elze}. 
So far, only a few models are known to be promising \cite{elze,hall1,buric}. 
As other related subjects, see Ref.\ \cite{ELF2}.

Recently, the present author proposed a new method to derive a model of QCH in a particle system \cite{koideqch}, 
based on the hypothetical view that quantization can be regarded as optimization of actions written by stochastic variables. 
The method is called stochastic variational method (SVM) \cite{yasue,koide-manual,zam}. 
Because a unified description of classical and quantum behaviors is possible in SVM \cite{koide-manual}, 
the derivation of QCH reduces to the problem of optimization with 
classical and stochastic variables.
At least a part of requirements in Ref.\ \cite{elze} are confirmed to be satisfied. 
The advantage of this approach is that conserved quantities are defined through 
the stochastic Noether theorem uniquely.

For practical applications of the QCH approach, it is important to consider systems 
with electromagnetic interactions. For this, we need to generalize the previous results 
to field theoretical systems.
In this paper, we study QCH in a simplified model of quantum electrodynamics (QED) 
where {\it classical} charged particles interacting with electromagnetic fields quantized in the Coulomb gauge condition. 
The precise meaning of the term ``classical" is defined in Sec.\ \ref{sec:qch}.

The QCH approach will be justified when the dynamical scales 
of classical and quantum degrees of freedom are clearly separated. 
In the present model, then, the behaviors of the {\it classical} charged particles are approximately given by adiabatic motions 
and a geometric phase such as Berry's phase will be observed. 
In fact, such a phase is discussed in another QCH model of a particle system and it is concluded that 
the usual form of Berry's phase is still reproduced \cite{zhang}. 
As we will see later, our geometric phase is slightly different from Berry's phase, 
which is, however, reproduced in the vanishing limit of the fluctuation of the charged particle trajectories.

\section{Quantization of a simplified QED model} \label{sec:2}

SVM was proposed by Yasue \cite{yasue} in 1981 so as to reformulate Nelson's stochastic quantization \cite{nelson}. 
As the pedagogical reviews, see Refs.\ \cite{koide-manual, zam}.
So far, this method has been applied mainly to quantum mechanical systems, but is possible to quantize field-theoretical systems \cite{koide-kk}.
Extending this result, we discuss the SVM quantization of a system with charged particles and electromagnetic fields in this section. 
Note that, however, the creation-annihilation effect of the charged particles is not taken into account in this model.
The QCH model of this system is discussed in Sec.\ \ref{sec:qch}.

\subsection{Setup}

In the SVM quantization, the origin of quantum fluctuations is attributed to stochastic motions 
of dynamical variables. 
Then, we need to introduce a stochastic trajectory $\hat{\bf r}(t)$ to represent a charged particle motion 
with a mass $M$. 
This stochastic variable is characterized by the following forward stochastic differential equation (SDE),
\begin{eqnarray}
\hspace*{-0.5cm} d\hat{\bf r}^{i} (t) = \frac{dt}{M}\left( {\bf p}^{i} (\hat{\bf r},\{ \hat{\bf A}\},t) - \frac{e}{c}\hat{\bf A}^{i} (\hat{\bf r},t) \right)
+ \sqrt{\frac{\hbar}{M}} d{\bf W}^{i} (t). \label{sder}
\end{eqnarray}
Here ${\bf W}(t)$ is a Wiener process, and $e$ and $c$ denote electric charge and speed of light, respectively.
The term ``forward" means the evolution forward in time, $dt > 0$.
The time dependence of the momentum ${\bf p}(\hat{\bf r},\{\hat{\bf A}\},t)$ is determined by employing 
a variational principle, and $\{\hat{\bf A} \}$ represents the configuration of a stochastic gauge field 
$\hat{\bf A}$. 
Note that Eq.\ (\ref{sder}) is reduced to the classical definition of velocity in the vanishing limit of $\hbar$.

To quantize electromagnetic fields in the Coulomb gauge condition, a similar SDE is employed for the transverse components of the gauge field, where 
$\{\hat{\bf A} \}$ satisfies $\nabla_{\bf x} \cdot \hat{\bf A}({\bf x},t) = 0$.  
Then the forward SDE for the transverse fields is given by  
\begin{eqnarray}
\hspace*{-0.7cm}d \hat{\bf A}^{i} ({\bf x}, t) = {\bf u}_{{\bf x}^{i}} (\hat{\bf r}, \{ \hat{\bf A} \},t)dt 
+ \sqrt{\frac{\hbar c^2}{(\Delta {x})^3}}\sum_j {\bf P}^{ij}_\bot d{\bf Y}^j_{\bf x} (t), \label{sdea}
\end{eqnarray}
where ${\bf P}_\bot$ is
\begin{equation}
{\bf P}^{ij}_\bot = \delta_{ij} - \nabla_{{\bf x}i}\Delta^{-1}_{\bf x}\nabla_{{\bf x}j}, \label{po}
\end{equation}
where $\Delta_{\bf x} = \sum_i \nabla^2_{{\bf x}i}$.
Again, ${\bf u}_{\bf x}$ is determined by the variation. 
Exactly speaking, $\hat{\bf A}$ is defined on a spatial lattice introduced by 
discretizing the space with $\Delta x$ being the grid interval. Then $\nabla_{\bf x}$ in Eq.\ (\ref{po}) and the Coulomb gauge condition 
is expressed as matrix. Note that we employ the periodic boundary conditions for $\hat{\bf A}$ 
on the lattice so that the partial integration formula for ${\bf x}$ is applicable \cite{koide-kk}.
The stochastic variable ${\bf Y}_{\bf x}(t)$ is a Wiener process at the lattice point ${\bf x}$ 
and there is no correlation with ${\bf Y}_{\bf x'}(t)$ when ${\bf x}\neq {\bf x}'$.
See Ref.\ \cite{koide-kk} for more details.

As is discussed in Refs.\ \cite{koide-manual, zam}, 
we need evolutions backward in time to define a variation fixing initial and final stochastic distributions.
Such backward evolutions are defined in a similar manner to the above forward equations, although we will not write down these explicitly.

Once the SDEs are given, we can define the probability distribution function of the particle position 
and the gauge field configuration as 
\begin{eqnarray}
\rho({\bf q}, \{{\bf a}\},t) = E[ \delta({\bf q}- \hat{\bf r}(t))\delta({\bf a}({\bf x})- \hat{\bf A}({\bf x},t)) ], \nonumber
\end{eqnarray}
where, $E[~~]$ denotes an expectation value for the Wiener processes. 
Initial distributions of 
the charged particle and the gauge field configuration are not shown in the above expression, but it does not 
affect the following discussions. See Ref.\ \cite{koide-manual} for more details.
Using the above SDEs for $\hat{\bf r}$ and $\hat{\bf A}$, the Fokker-Planck equation is obtained as  
\begin{eqnarray}
\partial_t \rho 
&=& - \frac{1}{M}\nabla_{\bf q} \cdot \left( {\bf p}_m({\bf q}, \{ {\bf a} \}, t) - \frac{e}{c}{\bf a}({\bf q}) \right) \rho  \nonumber \\
&& - \int d^3 {\bf x} \frac{\delta}{\delta {\bf a}({\bf x})} \cdot {\bf u}_{m{\bf x}}({\bf q}, \{{\bf a}\},t) \rho, \label{fpfull}
\end{eqnarray}
where $\delta/\delta {\bf a}({\bf x})$ is a functional derivative and 
\begin{eqnarray}
{\bf p}_m = \frac{{\bf p} + \tilde{\bf p}}{2} ,~~~~{\bf u}_{m{\bf x}} = \frac{{\bf u}_{\bf x} + \tilde{\bf u}_{\bf x}}{2}. 
\nonumber 
\end{eqnarray}
The symbol $\tilde{~~}$ represents quantities in backward SDEs which satisfy  
\begin{subequations}
\begin{eqnarray}
&& \hspace*{-1cm}\tilde{\bf p}({\bf q}, \{{\bf a}\},t) = {\bf p}({\bf q}, \{{\bf a}\},t) - \frac{\hbar}{2M} \nabla_{\bf q} \ln \rho ({\bf q}, \{{\bf a}\},t),
\label{ptil} \\
&& \hspace*{-1cm}\tilde{\bf u}_{\bf x} ({\bf q}, \{{\bf a}\},t) = {\bf u}_{\bf x} ({\bf q}, \{{\bf a}\},t) 
- \frac{\hbar c^2}{2} \frac{\delta \ln \rho ({\bf q}, \{{\bf a}\},t)}{\delta {\bf a}_\bot ({\bf x})} , \label{util}
\end{eqnarray}
\end{subequations}
where
\begin{eqnarray}
\frac{\delta}{\delta {\bf a}^i_\bot ({\bf x})} = \sum_j {\bf P}^{ij}_\bot \frac{\delta}{\delta {\bf a}^j({\bf x})}.
\end{eqnarray}
These are called consistency conditions and 
derived from the consistency of the two Fokker-Planck equations 
obtained from the forward and backward SDEs \cite{koide-manual}.

Note that stochastic trajectories are not smooth  
and the usual definition of the time derivative is not applicable. As was discussed by Nelson \cite{nelson}, instead,
the mean forward derivative $D$ and the mean backward derivative $\tilde{D}$ for the particle are introduced,  
\begin{eqnarray}
\hspace*{-0.5cm}
D \hat{\bf r} (t) &=& \lim_{dt \rightarrow 0+} E \left[ \frac{\hat{\bf r} (t + dt) - \hat{\bf r} (t)}{dt} \Big| {\cal P}_t \right], \nonumber \\
\hspace*{-0.5cm}
\tilde{D} \hat{\bf r}(t) &=& \lim_{dt \rightarrow 0-} E \left[ \frac{\hat{\bf r} (t + dt) - \hat{\bf r} (t)}{dt} \Big| {\cal F}_t \right], \nonumber 
\end{eqnarray}
respectively. 
These expectations are conditional averages, where ${\cal P}_t$ (${\cal F}_t$) indicates to fix 
$\hat{\bf r}(t')$ for $t'\le t$ $(t'\ge t)$. 
These are easily extended to field variables \cite{koide-kk}.
Using these definitions and Ito's lemma \cite{gardinar-book}, we obtain  
\begin{eqnarray}
&&\hspace*{-1.5cm} D f(\hat{\bf r}, \{ \hat{\bf A} \}, t)
=
\left[
\partial_t + \frac{1}{M}\left( {\bf p} \cdot \nabla_{\hat{\bf r}} - \frac{e}{c} \hat{\bf A} \right) 
+ \frac{\hbar}{2M} \Delta_{\hat{\bf r}} \right. \nonumber \\
&&\hspace*{-1.cm} \left. + \int d^3{\bf x} \left( {\bf u}_{\bf x} \cdot \frac{\delta}{\delta \hat{\bf A}({\bf x})}
+ \frac{\hbar c^2}{2} 
\frac{\delta^2}{\delta \hat{\bf A}^2_\bot ({\bf x})} \right)
\right] f(\hat{\bf r}, \{ \hat{\bf A} \}, t), \nonumber \\
&&\hspace*{-1.5cm} \tilde{D} f(\hat{\bf r}, \{ \hat{\bf A} \}, t)
=
\left[
\partial_t + \frac{1}{M}\left( \tilde{\bf p} \cdot \nabla_{\hat{\bf r}} - \frac{e}{c} \hat{\bf A} \right) 
- \frac{\hbar}{2M} \Delta_{\hat{\bf r}} \right. \nonumber \\
&&\hspace*{-1.cm} \left. + \int d^3{\bf x} \left( \tilde{\bf u}_{\bf x} \cdot \frac{\delta}{\delta \hat{\bf A}({\bf x})}
- \frac{\hbar c^2}{2} \frac{\delta^2}{\delta \hat{\bf A}^2_\bot({\bf x})} \right)
\right] f(\hat{\bf r}, \{ \hat{\bf A} \}, t), \nonumber 
\end{eqnarray}
where $f({\bf q},\{{\bf a}\},t)$ is a smooth functional.
Note that these are still stochastic quantities.

\subsection{Application}

Let us consider the quantization of a similar system with $N-$charged particles.  
The stochastic Lagrangian $L_{sto}$ is obtained by 
replacing the dynamical variables in the classical Lagrangian with the corresponding stochastic ones.
In the present case, we have 
\begin{eqnarray}
&& \hspace*{-1.8cm}L_{sto}
= 
\sum_{\alpha=1}^{N} \left[\frac{M_\alpha}{2}\frac{(D\hat{\bf r}_\alpha )^2 + (\tilde{D}\hat{\bf r}_\alpha )^2}{2} 
+ \frac{e_\alpha}{c} \frac{D\hat{\bf r}_\alpha + \tilde{D}\hat{\bf r}_\alpha}{2} \cdot \hat{\bf A} (\hat{\bf r}_\alpha,t) \right] \nonumber \\
&& \hspace*{-1cm} 
- \sum_{\alpha=1}^{N} e_\alpha A^0 (\hat{\bf r}_\alpha,t) 
-V(\{ \hat{\bf r} \}) 
+ \frac{1}{2} \int d^3 {\bf x} (\nabla_{\bf x} A^0 )^2 
\nonumber \\
&&\hspace*{-1.cm} +\frac{1}{2} \int d^3 {\bf x} 
\left[
\frac{1}{c^2} \frac{(D\hat{\bf A})^2 + (\tilde{D}\hat{\bf A})^2}{2}
- (\nabla_{\bf x} \times \hat{\bf A})^2 \right], \nonumber 
\end{eqnarray}
where $V(\{{\bf q}\})$ represents a potential with $\{ {\bf q} \}$ being ${\bf q}_1, {\bf q}_2, \cdots, {\bf q}_N$.
Note that $L_{sto}$ is reduced to the corresponding classical Lagrangian in the vanishing limit of $\hbar$.
As is well-known, there is an ambiguity for the replacement of the time derivative terms because we have now the two different time derivatives, 
$D$ and $\tilde{D}$. For the purpose of quantization, it is known that the time derivative terms should be replaced by the average of the two contributions 
as is done in the above.
See also Ref.\ \cite{koide1}.

The definitions developed in the previous subsection are easily extended to the system of $N-$charged particles.
By using them, we can calculate the variations of the stochastic variation of $L_{sto}$, 
leads to the stochastic Euler-Lagrange (EL) equations,
\begin{subequations}
\begin{eqnarray}
&& \hspace*{-2cm}\left[ \tilde{D} \frac{\partial L_{sto}}{\partial (D \hat{\bf r}^{i})} 
+ D \frac{\partial L_{sto}}{\partial (\tilde{D} \hat{\bf r}^{i})} - \frac{\partial L_{sto}}{\partial \hat{\bf r}^{i}} \right]
^{\hat{A}({\bf x},t) = {\bf a}({\bf x})}_{\hat{\bf r}_\alpha (t) = {\bf q}_\alpha} = 0, \label{eleq-r}\\
&& \hspace*{-2cm}\left[ \tilde{D} \frac{\partial L_{sto}}{\partial (D \hat{\bf A}^{i})} 
+ D \frac{\partial L_{sto}}{\partial (\tilde{D} \hat{\bf A}^{i})} 
+ \nabla_{\bf x}\cdot \frac{\partial L_{sto}}{\partial (\nabla_{\bf x} \hat{\bf A}^{i})} 
- \frac{\partial L_{sto}}{\partial \hat{\bf A}^{i}} \right]
^{\hat{A}({\bf x},t) = {\bf a}({\bf x})}_{\hat{\bf r}_\alpha (t) = {\bf q}_\alpha} \hspace{-1cm}= 0,  \nonumber \\
\label{eleq-a} \\
&& \hspace*{-2cm}\left[  
 \nabla_{\bf x} \cdot \frac{\partial L_{sto}}{\partial (\nabla_{\bf x} A^0)} 
- \frac{\partial L_{sto}}{\partial A^0} \right]
^{\hat{A}({\bf x},t) = {\bf a}({\bf x})}_{\hat{\bf r}_\alpha(t) = {\bf q}_\alpha} = 0. \label{eleq-a0}
\end{eqnarray}
\label{SEL}
\end{subequations}
Note that $\hat{\bf r}_\alpha(t)$ ($\hat{\bf A}({\bf x},t)$) is replaced by a time-independent variable ${\bf q}_\alpha$ (${\bf a}({\bf x})$) 
at the last of the calculations, 
because SVM requires that a stochastic action is optimized for any stochastic configuration.
The last equation is obtained by the variation of the c-number field $A^0$.

To simplify the results of the stochastic EL equations, 
we introduce a function $\theta$ as 
\begin{subequations}
\begin{eqnarray}
&& {\bf p}_{m\alpha} (\{{\bf q}\},\{{\bf a}\},t) = \hbar \nabla_{{\bf q}_\alpha} \theta (\{{\bf q}\},\{{\bf a}\},t), \\
&& {\bf u}_{m{\bf x}} (\{{\bf q}\},\{{\bf a}\},t) = \hbar c^2 \frac{\delta \theta (\{{\bf q}\},\{{\bf a}\},t)}{\delta {\bf a}_\bot ({\bf x})}.
\label{phase-u}
\end{eqnarray}
\end{subequations}
This phase $\theta$ does exist because Eqs.\ (\ref{eleq-r}) and (\ref{eleq-a}) lead to the common equation for $\theta$.
Note that $\theta$ becomes a multi-valued function when there are vortices in the geometric space (of ${\bf q}_\alpha$) 
and/or the functional space (of ${\bf a}({\bf x})$).

Then we can define the wave functional by 
\begin{eqnarray}
\Psi(\{{\bf q}\},\{ {\bf a} \},t) = \sqrt{\rho(\{{\bf q}\},\{ {\bf a} \},t)}e^{i\theta(\{{\bf q}\},\{ {\bf a} \},t)}, \nonumber 
\end{eqnarray}
leading to the probability distribution function $|\Psi|^2$.
The time evolutions of $\rho$ and $\theta$ are determined by Eqs.\ (\ref{fpfull}) and (\ref{eleq-r}) 
(or (\ref{eleq-a})), respectively, and  
can be unified in the form of the functional Schr\"{o}dinger equation,
\begin{eqnarray}
&&\hspace*{-1cm} i \hbar \partial_t \Psi
=
\left[ \sum_{\alpha=1}^N \left\{\frac{\left( -i\hbar \nabla_{{\bf q}_\alpha} 
- e_\alpha {\bf a}({\bf q}_\alpha)/c \right)^2}{2M_\alpha}
+ \frac{e_\alpha}{2} A^0 ({\bf q}_\alpha) 
\right\}
 \right. \nonumber \\
&&\hspace*{-0cm} \left. +  V(\{{\bf q}\}) + \frac{1}{2}\int d^3 {\bf x} \left\{ \mbox{\boldmath$\varepsilon$}^2 ({\bf x})   
+ {\bf b}^2 ({\bf x})   \right\}
\right] \Psi, \label{fse1}
\end{eqnarray}
where 
the magnetic operator is ${\bf b} ({\bf x}) = \nabla_{\bf x} \times {\bf a}({\bf x})$
and 
\begin{eqnarray}
\mbox{\boldmath$\varepsilon$} ({\bf x}) = i \hbar c \frac{\delta}{\delta {\bf a}_\bot ({\bf x})}.
\end{eqnarray}
The Coulomb potential $A^{0}({\bf q}_\alpha)$ is given by the solution of Eq.\ (\ref{eleq-a0}), 
\begin{eqnarray}
&& A^0 ({\bf q}_\alpha) = \frac{1}{4\pi} \sum_{\beta \neq \alpha} \frac{e_\beta}{|{\bf q}_\alpha - {\bf q}_\beta|}. \nonumber  
\end{eqnarray}
Note that (divergent) constant terms from the Coulomb self-energy 
are absorbed into the ambiguity for the phase of the wave functional.

The above functional representation of quantum field theory is already known and 
gives the same result as the operator representation \cite{huang}. 
In fact, our result reproduces the commutation relation in the Coulomb gauge condition \cite{weinberg,milonni}, 
\begin{eqnarray}
\hspace*{-0.5cm}  \left[ {\bf a}^{i}({\bf x}), \mbox{\boldmath$\varepsilon$}^j({\bf y}) \right] 
= i \hbar c \delta_{ij} \delta({\bf x} -{\bf y}) + i \hbar c\nabla_{{\bf x}i} \nabla_{{\bf x}j} \frac{1}{4\pi|{\bf x}-{\bf y}|}. \nonumber
\end{eqnarray}
Moreover the Hamiltonian operator in Eq.\ (\ref{fse1}) 
can be expressed in creation-annihilation operators \cite{koide-kk}.

\subsection{Ehrenfest's theorem}

Ehrenfest's theorem is satisfied in Eq.\ (\ref{fse1}).
 
The expectation values for the charged particle dynamics are given by  
\begin{eqnarray}
&& \hspace*{-1cm} \partial_{t} \langle {\bf q}_\alpha \rangle = \left\langle {\bf v}({\bf q}_\alpha ) \right\rangle, \nonumber \\
&& \hspace*{-1cm} \partial_t \langle -i\hbar \nabla_{{\bf q}_\alpha} \rangle = \frac{e}{c} \left\langle 
{\bf v}({\bf q}_\alpha ) \times {\bf b} ({\bf q}_\alpha) 
 + {\bf v}({\bf q}_\alpha ) \cdot \nabla_q {\bf a}({\bf q}) \right\rangle \nonumber \\
&& - \nabla_{q_\alpha}\langle  V (\{{\bf q}\}) + e_\alpha A^0 ({\bf q}_\alpha) \rangle \nonumber 
,
\end{eqnarray}
where 
\begin{eqnarray}
{\bf v}({\bf q}_\alpha) = \frac{1}{M_\alpha}(-i\hbar \nabla_{{\bf q}_\alpha} - \frac{e_\alpha}{c} {\bf a}({\bf q}_\alpha) ), \nonumber
\end{eqnarray}
and $\langle \ \ \rangle$ denotes the expectation value with $\Psi$.
These equations coincide with the corresponding classical equations besides the difference of c-numbers and operators.

On the other hand, the expectation values for the electromagnetic fields are 
\begin{eqnarray}
&& \hspace*{-1cm}  \nabla_{\bf x} \cdot \langle \mbox{\boldmath$\varepsilon$}({\bf x}) - \nabla_{\bf x} A^0({\bf x}) \rangle
= \rho ({\bf x},t), \nonumber \\
&& \hspace*{-1cm} \nabla_{\bf x} \cdot \langle {\bf b} ({\bf x}) \rangle = 0 , \nonumber \\
&& \hspace*{-1cm} \partial_t \langle {\bf b} ({\bf x}) \rangle
=  - c \nabla_{\bf x} \times \left\langle \mbox{\boldmath$\varepsilon$}({\bf x}) \right\rangle, \nonumber \\
&& \hspace*{-1cm}  \partial_t \left\langle \mbox{\boldmath$\varepsilon$}({\bf x}) \right\rangle
= c \nabla_{\bf x} \times \langle {\bf b} ({\bf x}) \rangle -\frac{1}{c} {\bf J}_\bot ({\bf x},t) , \nonumber 
\end{eqnarray}
where 
\begin{eqnarray}
&& \hspace*{-1cm}\rho ({\bf x},t) 
= 
\sum_\alpha \langle  e_\alpha \delta ({\bf x} - {\bf q}_\alpha) \rangle, \nonumber \\
&&\hspace*{-1cm} {\bf J}^{i}_\bot ({\bf x},t)
= \sum_\alpha \langle e_\alpha 
\delta({\bf x} - {\bf q}_\alpha) \sum_{j}{\bf P}^{ij}_\bot {\bf v}^j ({\bf q}_\alpha)
 \rangle. \nonumber 
\end{eqnarray}
Interpreting $\mbox{\boldmath$\varepsilon$}({\bf x})$ as a transverse electric operator, 
these coincide with Maxwell's equations.

\section{Quantum-classical hybrids} \label{sec:qch}

In this section, we derive a QCH model corresponding to Sec.\ \ref{sec:2} by extending the method in Ref.\ \cite{koideqch}.

\subsection{derivation}

Suppose that charged particles are approximately described as classical degrees of freedom under a certain initial condition 
and a parameter set of the system. 
The origin of quantum fluctuations in SVM is attributed to the fluctuation of dynamical variables induced by 
Wiener processes. 
Thus, instead of Eq.\ (\ref{sder}), we employ the following equation for a charged particle $\alpha$,
\begin{eqnarray}
d\hat{\bf r}_\alpha (t)  = \frac{dt}{M_\alpha}\left( {\bf p}_\alpha (\{\hat{\bf r}\},\{ \hat{\bf A} \},t) - \frac{e_\alpha}{c}\hat{\bf A} (\hat{\bf r}_\alpha,t) \right). \label{app-qch}
\end{eqnarray}
This is the {\it classicalization} procedure of the particle degrees of freedom to obtain QCH. 
Note that $\hat{\bf r}_\alpha (t)$ still fluctuates through the $\hat{\bf A}$ dependence.
This modification immediately affects the calculation of the mean derivatives,
\begin{eqnarray}
\hspace*{-0.5cm} M_\alpha D \hat{\bf r}_\alpha (t) = M_\alpha \tilde{D} \hat{\bf r}_\alpha (t) = 
{\bf p}_\alpha (\{\hat{\bf r}\},\{ \hat{\bf A} \},t) - \frac{e_\alpha}{c}\hat{\bf A} (\hat{\bf r}_\alpha,t). \nonumber 
\end{eqnarray}

Let us calculate the stochastic variation using Eq.\ (\ref{app-qch}) for $\hat{\bf r}_\alpha$ and Eq.\ (\ref{sdea}) for $\hat{\bf A}$.
The stochastic EL equations (\ref{eleq-r}), (\ref{eleq-a}) and (\ref{eleq-a0}) 
are still formally satisfied, but $\hat{\bf r}_\alpha$ is replaced not by ${\bf q}_\alpha$ 
but by ${\bf f}_\alpha (\{{\bf a}\},t)$ which is defined soon later.
This is because SVM requires that the action is optimized for any configuration of 
$\hat{\bf A} ({\bf x}, t)$ and thus it is replaced by the time-independent function ${\bf a}({\bf x})$. 
However, $\hat{\bf r}_\alpha (t)$ now fluctuates only through the $\hat{\bf A}({\bf x}, t)$ dependence because of the above classicalization. 
Once $\hat{\bf A} ({\bf x}, t)$ is replaced by ${\bf a}({\bf x})$, $\hat{\bf r}_\alpha (t)$ does not fluctuate any more 
and hence it cannot be replaced simply by ${\bf q}_\alpha$.
The right hand side of Eq.\ (\ref{app-qch}) then  
becomes a function not only of $t$ but also of ${\bf a}({\bf x})$. 
Therefore we introduce a new quantity defined by the solution of the following equation,
\begin{eqnarray}
\partial_t {\bf f}_\alpha (\{{\bf a}\},t) = \frac{1}{M_\alpha}\left( {\bf p}_\alpha (\{{\bf f}\},\{ {\bf a} \},t) 
- \frac{e}{c}{\bf a} ({\bf f}_\alpha) \right). \label{pse-tra}
\end{eqnarray}
We call ${\bf f}_\alpha (\{{\bf a}\},t)$ quasi trajectory function, which  
is substituted into $\hat{\bf r} (t)$ in the stochastic EL equations.

In our QCH, we need the probability distribution function of ${\bf a}({\bf x})$ when the charged particles are found 
at $\{ {\bf f} \}$,
\begin{eqnarray}
\hspace*{-1cm}\rho_{a}(\{ {\bf a} \},t) = \int d^3 {\bf x} \prod_{\alpha}d^3 {\bf q}_\alpha 
 E[\delta({\bf q}_\alpha - {\bf f}_\alpha )\delta({\bf a}({\bf x}) - \hat{\bf A}({\bf x},t)) ]. \nonumber 
\end{eqnarray}
Two different Fokker-Planck equations are obtained by substituting, respectively, 
the solutions of the forward and backward SDEs of $\hat{\bf A}({\bf x},t)$ into $\rho_a$, 
but both should be equivalent. Then we obtain a new consistency condition   
instead of Eq.\ (\ref{util}), 
\begin{equation}
\hspace*{-0.5cm}\tilde{\bf u}_{\bf x} (\{{\bf f}\}, \{{\bf a}\},t) = {\bf u}_{\bf x} (\{{\bf f}\}, \{{\bf a}\},t) 
- \frac{\hbar c^2}{2} \frac{\delta \ln \rho_a (\{{\bf a}\},t)}{\delta {\bf a}_\bot ({\bf x})}.
\end{equation}

Using this, the stochastic EL equation for the charged particles is calculated as
\begin{eqnarray}
&&\hspace*{-1.5cm} \left( \partial_t + \sum_\alpha {\bf v}^{qch}_\alpha \cdot \nabla_{{\bf f}_\alpha}
+ \hbar c^2 \int d^3 {\bf y}
\frac{\delta \theta}{\delta {\bf a}_\bot ({\bf y})}
\cdot\frac{\delta }{\delta {\bf a}_\bot ({\bf y})}
\right)  {\bf v}^{qch}_\alpha  \nonumber \\
&&\hspace*{-1cm} = - \frac{1}{M_\alpha}\nabla_{{\bf f}_\alpha} V(\{{\bf f}\}) \nonumber \\
&&\hspace*{-1cm} + \frac{e_\alpha}{cM_\alpha} {\bf v}^{qch}_\alpha  \times {\bf b}({\bf f}_\alpha)
+ \frac{e_\alpha}{M_\alpha} (  \mbox{\boldmath$\varepsilon$}({\bf f}_\alpha)- \nabla_{{\bf f}_\alpha} A^0 ({\bf f}_\alpha)), \label{qch-cl}
\end{eqnarray}
where the phase $\theta$ is defined in the same manner as before, and 
\begin{eqnarray}
&& \hspace*{-1cm}{\bf v}^{qch}_\alpha (\{{\bf f}\},\{{\bf a}\},t) 
= \frac{\hbar}{M_\alpha} \nabla_{{\bf f}_\alpha} \theta (\{{\bf f}\},\{{\bf a}\},t)  - \frac{e_\alpha}{cM_\alpha} {\bf a}({\bf f}_\alpha) , \nonumber \\
&& \hspace*{-1cm} A^0 ({\bf f}_\alpha) 
= \frac{1}{4\pi} \sum_{\beta \neq \alpha} \frac{e_\beta}{|{\bf f}_\alpha - {\bf f}_\beta|}. \nonumber
\end{eqnarray}

On the other hand, the dynamics of the quantized gauge field is described by the functional Schr\"{o}dinger equation,  
\begin{eqnarray}
\hspace*{-0.5cm} i \hbar \partial_t \Psi_a ({\bf f},\{ {\bf a}_\bot\},t) 
 = 
{\cal H} (\{{\bf v}^{qch},{\bf f}\})  \Psi_a ({\bf f},\{ {\bf a}_\bot\},t),
\label{qch-qm}
\end{eqnarray}
where
\begin{eqnarray}
&&\hspace*{-1.2cm} {\cal H} (\{{\bf v}^{qch},{\bf f}\}) 
= \sum_\alpha \left\{ \frac{M_\alpha ({\bf v}^{qch}_\alpha)^2}{2} + \frac{e_\alpha}{2}A^0({\bf f}_\alpha) \right\}
+V(\{{\bf f}\})  \nonumber \\
&& +
\frac{1}{2}\int d^3 {\bf x} 
\left\{ \mbox{\boldmath$\varepsilon$}^2({\bf x}) + {\bf b}^2({\bf x}) \right\}. \label{hameff}
\end{eqnarray}
Here the wave functional is defined by 
\begin{eqnarray}
\Psi_a(\{{\bf f}\},\{ {\bf a} \},t) = \sqrt{\rho_a(\{ {\bf a} \},t)}e^{i\theta(\{{\bf f}\},\{ {\bf a} \},t)}, \nonumber 
\end{eqnarray}
and thus the probability distribution function of the gauge field configuration is expressed as $|\Psi_a|^2$.
Note that the time derivative appearing on the left hand side is the partial time derivative $\partial_t$. 
If it is expressed with the total time derivative $d/dt$, the equation is reexpressed as 
Eq.\ (\ref{fse-tt}).

\subsection{properties of QCH}

In the following, we investigate the properties of our QCH model. 

\subsubsection{no interaction limit}\label{noint}

When there is no interaction between classical and quantum degrees of freedom, $e_\alpha = 0$, 
as is seen from Eq.\ (\ref{qch-cl}), the velocity ${\bf v}^{qch}$ 
can be independent of ${\bf a}({\bf x})$.
Then the quasi trajectory function ${\bf f}$ has a definite path and Eq.\ (\ref{qch-cl}) 
coincides with Newton's equation of motion.
On the other hand, ${\bf v}^{qch}$ and ${\bf f}$ in the functional Schr\"{o}dinger equation 
become only functions of time 
and such terms are absorbed into the ambiguity of the phase of the wave functional, 
leading to ${\cal H}  = \frac{1}{2}\int d^3 {\bf x} 
\left\{ \mbox{\boldmath$\varepsilon$}^2({\bf x}) + {\bf b}^2({\bf x}) \right\}$.

That is, our model is separated into two decoupled equations in this limit as is expected: 
Newton's equation of motion and the functional Schr\"{o}dinger equation.

\subsubsection{energy conservation}

The conserved quantities of this model are  
defined by applying the stochastic Noether theorem \cite{koideqch,koide-kk,misawa}.
Then the conserved energy, associated with the time-translation invariance of the stochastic action, 
is given by  
\begin{eqnarray}
E = \langle {\cal H}(\{{\bf v}^{qch},{\bf f}\}) \rangle_a, \nonumber 
\end{eqnarray}
where ${\cal H}$ is defined by Eq.\ (\ref{hameff}), 
and $\langle ~~ \rangle_a$ denotes the expectation value with the wave functional $\Psi_a$.
In fact, we can directly confirm that 
\begin{eqnarray}
 \frac{dE}{dt} = \int [D{\bf a}] (\partial_t + \sum_\alpha {\bf v}^{qch}_\alpha \cdot \nabla_{{\bf f}_\alpha}) \Psi^* 
{\cal H}\Psi = 0. \nonumber 
\end{eqnarray}

\subsubsection{Extended Ehrenfest theorem}

The expectation values of the charged particle dynamics are calculated as 
\begin{eqnarray}
&&\hspace*{-1cm} \frac{d}{dt}E [\hat{\bf r}_\alpha] =  \langle {\bf v}^{qch}_\alpha \rangle_a, \nonumber \\
&&\hspace*{-1cm} \frac{d}{dt} \langle {\bf v}^{qch}_\alpha \rangle_a
=
 - \frac{1}{M_\alpha} \langle \nabla_{{\bf f}_\alpha} V(\{{\bf f}\}) \rangle_a \nonumber \\
&&\hspace*{-1cm} +  \frac{e_\alpha}{cM_\alpha} \langle {\bf v}^{qch}_\alpha  \times {\bf b}({\bf f}_\alpha) \rangle_a
+ \frac{e_\alpha}{M_\alpha} \langle   \mbox{\boldmath$\varepsilon$}({\bf f}_\alpha)- \nabla_{{\bf f}_\alpha} A^0 ({\bf f}_\alpha) \rangle_a . \nonumber 
\end{eqnarray}
The first equation is obtained from Eq.\ (\ref{app-qch}).
These are the essentially same results obtained in the previous work (Eqs.\ (18) and (19) in Ref.\ \cite{koideqch}) and  
reproduce the corresponding classical equations of motion, ignoring the difference between $\hat{\bf r}_\alpha$ and ${\bf f}_\alpha$.
In fact, it is worth mentioning that the difference between $E [\hat{\bf r}_\alpha]$ and $\langle {\bf f}_\alpha \rangle_a$ 
plays a role of an order parameter to characterize the limitation of the QCH approach \cite{koideqch}.

Similarly, the equations of the electromagnetic fields are 
\begin{subequations}
\begin{eqnarray}
&&\hspace*{-1cm} \nabla_{\bf x} \cdot \langle \mbox{\boldmath$\varepsilon$}({\bf x}) - \nabla_{\bf x} A^0 ({\bf x})\rangle_a = 
\rho^{qch} ({\bf x},t),  \\
&& \hspace*{-1cm} \nabla_{\bf x} \cdot \langle {\bf b} ({\bf x}) \rangle_a = 0 ,  \\
&& \hspace*{-1cm} \frac{d}{dt} \langle {\bf b} ({\bf x}) \rangle_a
=  - c \nabla_{\bf x} \times \left\langle \mbox{\boldmath$\varepsilon$}({\bf x}) \right\rangle_a,  \\ 
&&\hspace*{-1cm} \frac{d}{dt} \langle \mbox{\boldmath$\varepsilon$} ({\bf x}) \rangle_a 
= c \nabla_{\bf x} \times \langle {\bf b} ({\bf x}) \rangle_a - \frac{1}{c}{\bf J}^{qch}_\bot({\bf x},t), \label{ehren1} 
\end{eqnarray}
\end{subequations}
where 
\begin{subequations}
\begin{eqnarray}
&&\hspace*{-1cm} \rho^{qch} ({\bf x},t) = \sum_\alpha \langle  e_\alpha \delta ({\bf x} - {\bf f}_\alpha) \rangle_a , \\
&&\hspace*{-1cm} ({\bf J}^{qch}_\bot)^{i} ({\bf x},t)
= \sum_\alpha \langle e_\alpha 
\delta({\bf x} - {\bf q}_\alpha) \sum_{j}{\bf P}^{ij}_\bot ({\bf v}^{qch}_\alpha)^j 
 \rangle_a  \nonumber \\
&& + \sum_\alpha \langle ({\bf v}^{qch}_\alpha \cdot \nabla_{{\bf f}_\alpha} {\bf u}_{m{\bf x}}) \rangle_a . \label{currentnew}
\end{eqnarray}
\end{subequations}
The first three equations reproduce the corresponding classical ones, but 
the current in Eq.\ (\ref{ehren1}) is modified. 
The second term in Eq.\ (\ref{currentnew}) represents a new contribution corresponding to a kind of the displacement current. 
In fact, the electric field is given by 
$\langle \mbox{\boldmath$\varepsilon$}({\bf x}) \rangle_a = -\left\langle  
{\bf u}_{m{\bf x}} / c \right\rangle_a$, and ${\bf u}_{m{\bf x}}$ 
depends on time even through the $\{{\bf f}\}$ dependence, and 
the second term in ${\bf J}^{qch}_\bot$ is attributed to this time derivative in 
$\langle \mbox{\boldmath$\varepsilon$}({\bf x}) \rangle_a$. 
Note that, even if such a correction is added, the equation of continuity for the charge density and the conducting current 
is still hold.
Moreover, this correction term disappears in the vanishing limit of $e_\alpha$ as was discussed in Sec.\ \ref{noint}.

\section{Geometric phase} \label{sec:gp}

To apply the QCH approach successfully, 
the time scales of classical and quantum degrees of freedom will be clearly separated so that we can employ the adiabatic approximation.
Then a geometric phase will appear in the QCH wave functional $\Psi_a$ 
as in quantum mechanics. 
As a recent review of the geometric phase, 
see Ref.\ \cite{gp-review}. 

To discuss the geometric phase induced by the charged particle trajectories, 
let us introduce eigenstates by 
\begin{eqnarray}
{\cal H}(\{{\bf v}^{qch},{\bf f}\}) \phi_n  (\{{\bf a}\},t) = E_n (t)\phi_n (\{{\bf a}\},t), \nonumber 
\end{eqnarray}
and we assume that $\phi_n$ forms the complete orthonormal set.
We further consider that $E_n(t)$ is discretized and there is no degeneracy. 
See also discussion in Sec.\ \ref{sec:cr}.

From Eq.\ (\ref{qch-qm}), the total time derivative of $\Psi_a$ is 
\begin{eqnarray}
i \hbar \frac{d}{dt} \Psi_a = i \hbar ( \partial_t + \sum_\alpha {\bf v}^{qch}_\alpha \cdot \nabla_{{\bf f}_\alpha} ) \Psi_a
= [{\cal H} + {\cal H}_\delta ] \Psi_a, \label{fse-tt}
\end{eqnarray}
where ${\cal H}_\delta$ 
is a function of ${\bf v}^{qch}$ and disappears for ${\bf v}^{qch}=0$.

Let us expand $\Psi_a$ by the above eigenstates, 
\begin{eqnarray}
\Psi_a (\{{\bf f}\},\{{\bf a}\},t)= \sum_l c_{nl}(t) e^{-iF_l(t)/\hbar} \phi_l (\{{\bf a}\},t). \nonumber 
\end{eqnarray}
As an initial condition, we consider 
$\Psi_a (\{{\bf f}\},\{{\bf a}\},0) = \phi_n (\{{\bf a}\},0)$, leading to $c_{nl}(0) = \delta_{nl}$. 
Substituting into Eq.\ (\ref{fse-tt}), we obtain two equations. 
One is for $c_{nl}(t)$, 
\begin{eqnarray}
&&\hspace*{-1cm} \frac{d}{dt}c_{nl}(t)   
+ \sum_{m\neq l} \frac{c_{nm}(t)e^{-\frac{i}{\hbar}(F_m(t)-F_l(t))}}{E_m - E_l} \langle l, t | \dot{\cal H} | m, t \rangle_0  
 \nonumber \\
&&\hspace*{-1cm} = 
-\frac{i}{\hbar} \left\{ 
\sum_{m\neq l} c_{nm}(t) 
e^{-\frac{i}{\hbar} (F_m(t)-F_l(t))}
\langle l, t | {\cal H}_\delta | m, t \rangle_0 
\right\}, \nonumber 
\end{eqnarray}
where $\langle l, t |\ \ | m, t \rangle_0$ denotes the expectation value with 
$\phi_l^*$ and $\phi_m$.
In the adiabatic limit where the motions of charged particles are very slow, ${\bf v}^{qch} \sim 0$, 
the solution of this equation is $c_{nl} (t) = \delta_{nl}$. 
This is the well-known result in quantum mechanics and thus 
any dynamical change of $\Psi_a$ is absorbed into the behavior of $F_n(t)$.

The other equation is for $F_n(t)$, 
\begin{eqnarray}
\frac{d}{dt} {F}_n (t) 
=
\langle n, t | {\cal H} + {\cal H}_\delta | n, t \rangle_0 
- i \hbar  \langle n,t | \frac{d}{dt} |n, t\rangle_0 . \nonumber 
\end{eqnarray}
Because ${\cal H}$ depends on time through ${\bf f}$ and ${\bf v}^{qch}$, the time derivative of $|n, t\rangle_0$ 
in the second term is given by the derivatives for those quantities. 
Moreover, $d{\bf v}^{qch}/dt$ will be sufficiently small in the adiabatic process. 
Then this phase is given by 
\begin{eqnarray}
F_n (T) = \int^T_0 dt \langle n, t | {\cal H} + {\cal H}_\delta | n, t \rangle_0 - \hbar \gamma_n (t),\nonumber
\end{eqnarray}
where 
\begin{eqnarray}
\gamma_n (t) = i \sum_\alpha \langle n,t | \int_{C_\alpha(\{{\bf a}\})} d {\bf f}_\alpha\cdot \nabla_{{\bf f}_\alpha} |n, t\rangle_0. 
\label{berry} 
\end{eqnarray}
The path along the trajectory ${\bf f}_\alpha$ is given by $C_\alpha(\{{\bf a}\})$ 
for a given gauge field configuration $\{{\bf a}\}$.
Equation\ (\ref{berry}) is the geometric phase corresponding to Berry's phase, but the path integration 
now depends on $\{{\bf a}\}$.
The usual expression of Berry's phase is reproduced in the vanishing limit of this fluctuation 
of the charged particle trajectories.

This geometric phase is essentially calculated from the phase of $|n, t\rangle_0$. 
When ${\bf f}_\alpha$ forms a closed path, this integration can be calculated with Stokes' theorem.
Then, for $\gamma_n (t)$ to be finite, the phase should be a multi-valued function for the ${\bf f}_\alpha$ dependence.
See also the argument below Eq.\ (\ref{phase-u}).

\section{Discussions and concluding remarks} \label{sec:cr}

In this paper, we derived a model of QCH for a system where {\it classical} charged particles 
interact with electromagnetic fields quantized in the Coulomb gauge condition. 
This formulation is based on a hypothetical perspective for quantization in SVM, 
and the present result was obtained by extending the method in Ref.\ \cite{koideqch} 
to a field-theoretical system. Confirming the consistency of the derived model, we found that the 
quantum-classical interactions induce a displacement current and then Ehrenfest's theorem is satisfied in a modified manner.
For displacement currents appearing in various quantum transports, see Ref.\ \cite{displace}.

This is the derivation in the Coulomb gauge condition and 
thus it is interesting to ask whether a gauge-invariant formulation is possible, 
although the SVM quantizations in other gauge conditions are not yet understood.
Moreover, we did not consider the rotation of charged particles in this QCH model. 
It was found recently that the classical Maxwell-Lorentz equation with rotating charge can be formulated in 
the form of the variational principle \cite{imay}. To apply SVM to this Lagrangian, 
we need to generalize the framework of SVM so as to describe a stochastic rotation.
This problem will be related even to the description of the spin degree of freedom \cite{holland}. 
Such a generalization is interesting but left as a future task.

We further studied the geometric phase induced by the adiabatic motions of the charged particles, which 
appears in the wave functional of the gauge field configuration. 
The obtained phase is very similar to Berry's phase, but 
the path integration depends on the gauge field configuration $\{{\bf a}\}$.
This path fluctuation is attributed to the backreaction effect of interactions 
between classical and quantum degrees of freedom which is not considered in the usual Berry's phase, and 
thus this modification is reasonable.
The geometric phase is already investigated in a different QCH model of a particle system  \cite{zhang}, 
and it is claimed that the usual Berry's phase is still reproduced.
This difference will come from the different backreaction mechanism in QCH models.

Berry's phase is usually discussed for particle dynamics, 
but such a geometric phase exists even in electromagnetism. 
Such works focus mainly on a phase induced by the trajectory of photons \cite{phase-photon}. 
A situation analogous to our case is considered in Ref.\ \cite{carollo}, where 
a feasible experiment to observe a geometric phase in the wave functional (Fock space vector) is proposed.

In the derivation, we assumed that there is no degeneracy in the energy eigenstate. 
Although the spatial isotropy is broken due to charged particle trajectories, 
there may still exist some cases with degeneracy.  
It is known that the non-Abelian generalization of Berry's phase is observed by considering the effect of degeneracy \cite{wil}.

The Aharonov-Bohm effect is another interesting phenomenon \cite{ab-review,holland}. 
To discuss this effect, it will be more appropriate to consider another QCH limit where 
the gauge field is substituted by {\it classical} degrees of freedom. In Ref.\ \cite{boyer}, 
the Aharonov-Bohm effect is analyzed in classical electromagnetism and thus it is interesting to study 
how the result is modified by introducing quantum effects step-by-step in QCH.

So far, we have focused on the quantum mechanical aspect of SVM, but the applicability 
of SVM is more general. 
We can formulate a variational principle even for dissipative systems (Navier-Stokes-Fourier equation) 
and coarse-grained dynamics (Gross-Pitaevskii equation) \cite{koide1}. 
Moreover, it has not yet been understood what is a more precise and general  
quantization procedure \cite{gazeau}. The SVM quantization provides 
an interesting perspective on this problem.

The author acknowledges C.\ M.\ S. da Concei\c{c}\~{a}o, J.\ P.\ Gazeau, T.\ Kodama, K.\ Tsushima and I.\ A.\ Reyes for fruitful discussions and comments.
This work is financially supported by CNPq.

\end{document}